\documentstyle[twocolumn,aps,prb,epsfig,floats]{revtex}
\begin{document}
\draft
\title{Mesoscopic decoherence in Aharonov-Bohm rings}

\author{A.E.\ Hansen, A.\ Kristensen, S.\ Pedersen, C.B.\ S\o rensen, 
and P.E.\ Lindelof}
\address{The Niels Bohr Institute, University of Copenhagen, Universitetsparken 
5, DK-2100 Copenhagen, Denmark}
\date{\today}
\maketitle

\begin{abstract}
We study electron decoherence
by measuring the temperature dependence of Aharonov-Bohm (AB) oscillations in
quasi-1D rings, etched in a high-mobility GaAs/GaAlAs heterostructure.
The oscillation amplitude is influenced both by phase-breaking and by 
thermal averaging.
%which occurs in a conductance measurement at finite temperature.
Thermal averaging is important when the temperature approaches the energy scale,
on which the AB oscillations shift their phase.
For the phase-breaking,
it is  demonstrated that the damping of the
oscillation amplitude is proportional to
the length of the interfering paths.
For temperatures $T$ from 0.3 to 4 $\rm K$ we find 
the phase coherence length $\it L_{\phi}$ 
$\propto$ $T^{-1}$, 
close to what has been
reported for open quantum dots.
This might indicate that the $T^{-1}$ decoherence rate is a general
property of open and ballistic mesoscopic systems.
\end{abstract}

The understanding of decoherence
in quantum mechanical systems gives valuable insight into
the cross-over from quantum to classical behavior.
Quantum phenomena like weak localization, universal
conductance fluctuations and the Aharonov-Bohm effect, that are observed
in mesoscopic electronic systems, make these systems well suited for studying
decoherence.
The loss of electron phase coherence 
is interesting in its own right because it
reveals information about the 
fundamental physics of the electron scattering mechanisms.
Moreover, from the perspective of possible
phase-coherent mesoscopic electronic devices
\cite{devices},
knowledge of phase-breaking length and time scales is crucial.

At low temperatures, electron-electron scattering is usually the dominating source
of phase-breaking.
In disordered 1D and 2D conductors, the loss of phase coherence at low 
temperatures has been studied intensively, both theoretically 
and experimentally \cite{disordered}.
In clean electron systems, the number of investigations are fewer
\cite{Murphy95,Yacoby9195,Bird95,Huibers98}.
In 2D, experiments consistent with the expected \cite{Smith} 
electron-electron scattering time
$\tau_{\phi} \sim (T^{2} \ln T)^{-1}$ 
has been carried out \cite{Murphy95,Yacoby9195}.
In open quantum dots (a 0D system),
an unexpected $T^{-1}$ contribution was found
\cite{Bird95,Huibers98}.
In general, phase breaking mechanisms in ballistic, mesoscopic systems of 
dimensionality less than 2, are presently not well understood.

Aharonov-Bohm (AB) rings are obvious systems for probing phase coherence.
Here, the interference of two electron paths 
leads to conductance oscillations of period $h/e$ [or frequency $e/h$] in the magnetic 
flux enclosed by the paths. The oscillation amplitude is a direct measure of the
interference strength, and it has been used to study decoherence in disordered
systems \cite{disorderedRings}.
For AB rings with a 2DEG elastic mean free 
path longer than the circumference of the device 
(e.g.\ \cite{Ford90,Liu94,Mailly97,Pedersen00,Leo00,Yang00,Casse} 
and references 
therein), systematic studies
of phase-breaking have been scarce.
%To our knowledge, only Ref.\ \cite{Casse} directly addresses
%the temperature dependence of the AB oscillations. 
%In this work, however, the effect of phase breaking was considered to be
%small. 

In this paper, we report measurements of the phase coherence length 
$\it L_{\phi}$ via the temperature dependence of
AB conductance oscillations in quasi-1D rings,
made by shallow etching in GaAs/GaAlAs heterostructures.
Two mechanisms are important for the temperature dependence
of the oscillation amplitude:
phase-breaking, and thermal averaging. 
At finite temperature,
the measured 
conductance is a weighted average over an energy interval of finite width, proportional
to the temperature. We discuss, how
thermal averaging influences the AB oscillation amplitude through the phase changes of
the oscillations. In the experiment,
we detect AB oscillations due to the interference of electrons
that encircle the ring up to $n$ $=$ 6 times.
This is observed as
peaks in the Fourier spectra of the magnetoconductance at
multiples $ne/h$ of the fundamental AB frequency.
Accounting for the effect of thermal averaging, 
we find that the damping of the amplitude due to phase-breaking
depends linearly on $n$, showing directly the relaxation nature of the decoherence.
We find the phase coherence length $\it L_{\phi}$ $\propto$ $T^{-1}$.

\begin{figure}
\centerline{
\includegraphics[angle=0, width=8.5cm]{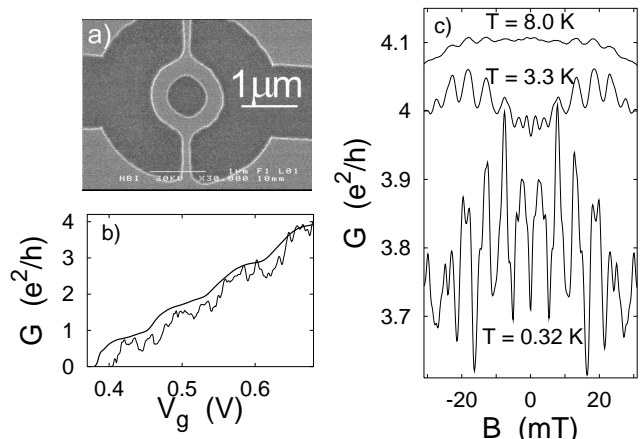}
}
\caption{
a) SEM image of the ring before gate deposition. 
In the dark areas the donor layer is etched away.
The quantum wires defining the arms  of the ring
are etched 280 $\rm nm$ wide, while the wires connecting the ring to the 2DEG 
reservoirs are 100 $\rm nm$ wide.
b) Conductance $G$ vs.\ gate voltage $V_g$ at $T$ $=$ 5.8 $\rm K$ (thick line),
and  0.32 $\rm K$ (thin line).
c) Examples of AB oscillations measured at fixed gate voltage
and different temperatures. Traces at high temperatures are offset for clarity.}
\label{abtempSEM_1}
\end{figure}

The AB rings are fabricated in a two-dimensional electron gas (2DEG) 
situated 90 $\rm nm$ below the surface of 
a modulation doped  GaAs/GaAlAs heterostructure.
At liquid He temperatures, the unpatterned
2DEG density and mobility is $n$ = 2.0$\times$10$\rm ^{15}$ m$\rm ^{-2}$
and $\mu$ = 80 m$\rm ^2/Vs$, corresponding to an
elastic mean free path $l_e$ $=$  6 
$\mu \rm m$. The lateral confinement is obtained by a shallow wet-etch, 
and the device is covered by a metal gate electrode.
Details on the sample fabrication have been presented elsewhere \cite{Pedersen00}.
The sample was cooled in a $\rm ^{3}$He cryostat,
and the  conductance was measured in a two-terminal configuration.
We used a conventional voltage biased
lock-in technique, with an excitation voltage of 31.6 $\mu \rm V$ oscillating 
at a frequency of 116.5 $\rm Hz$. 

Two samples with identical designs have been investigated in detail.
Here we present measurements on one of them. The main results are 
reproduced in the second sample.
A Scanning Electron Microscope (SEM) image of the ring is shown
in Fig.\ \ref{abtempSEM_1}a. The ring has a circumference of
3 $\mu \rm m$, $<$ $l_e$.
The gate voltage dependence of the conductance is shown in Fig.\
\ref{abtempSEM_1}b. At $T$ $=$ 5.8 $\rm K$, the conductance increases in steps
of height $\sim$ 
$e^2/h$, due to the conductance quantization in the narrow exit
and entrance wires \cite{Pedersen00}.
The steps 
should not be taken as a sign of the population of transverse subbands
in the ring itself, where the wires defining the arms are wider \cite{dE}. 
At $T$ $=$ 0.3 $\rm K$, a UCF-type signal is superimposed on the steps, indicating
a large degree of phase coherence in the ring at this temperature.

In Fig.\ \ref{abtempSEM_1}c we show an example of the temperature evolution of the
conductance in a perpendicular magnetic field $B$. 
The AB oscillation amplitude decreases with temperature.
The period of the oscillations is  5.4 $\rm mT$, in agreement with the 
period $h/(e \pi r^2)$ $=$ 5.5 $\rm mT$
calculated from the average radius $r$  $=$ 490 $\rm nm$ 
of the ring. 

\begin{figure}
\centerline{
\includegraphics[angle=0, width=8.5cm]{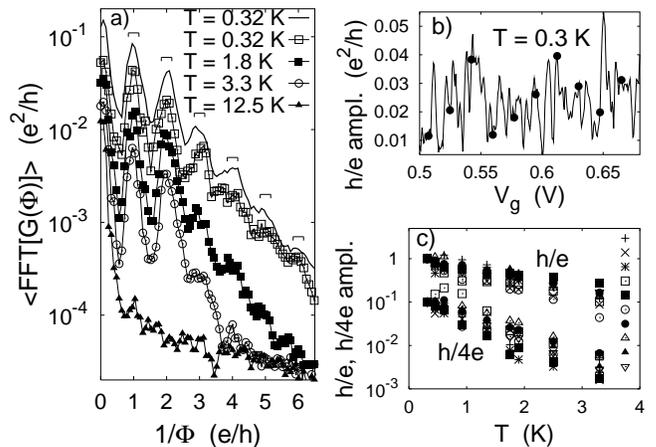}
}
\caption{
a) Fourier spectra at different temperatures, given in the legend,
as function of inverse magnetic flux, $\Phi=B\pi r^2$ through the ring.
The spectra
have been averaged for 10 different gate voltages. The spectrum 
marked with a thick line results from an
average of 500 gate voltages and is offset vertically for clarity.
The symbols mark the intervals in which the
Fourier coefficients are integrated to calculate the oscillation strengths.
b) Amplitude of $h/e$ oscillations as function of gate voltage. The
filled circles mark the 10 gate voltages used in the further analysis.
c) Amplitude of $h/e$ and $h/4e$ oscillations for 10 different gate voltages
$V_g$, normalized to 1 and 0.1 at $T$ $=$ 0.3 $\rm K$, as function of temperature.
The gate voltage increases from $V_g$ $=$ 0.51 $V$ to 0.67 $V$ in order of the
symbols plotted to the right.
}
\label{abtempFFT_5}
\end{figure}

To quantify the degree of coherence in the ring from the 
AB oscillations, we
compute the Fast Fourier Transform (FFT) of the conductance  $G(B)$, measured
in the magnetic field interval as shown in Fig.\ \ref{abtempSEM_1}c.
A measure of the AB oscillation amplitude is obtained by integrating the 
$ne/h$ FFT peaks
in the intervals as indicated in Fig.\ \ref{abtempFFT_5}a.
The amplitudes vary  with gate voltage, as exemplified for the $e/h$ peak in
Fig.\ \ref{abtempFFT_5}b.
It %of the AB oscillations 
is sensitive
to the phase difference $\Delta(k_F L)$ at zero magnetic field between electron paths
in either arm of the device \cite{Buttiker,Pedersen00}. The typical phase pickup
in an arm is $\it k_F L$ $\sim$ 200, 
where $\it L$ $=$ 1.5 $\rm \mu m$ is half the circumference
of the ring and $k_F$ is the Fermi wave number $\sim$ 1.5$\times$10$^8$ $\rm m^{-1}$.
%A maximum change in amplitude
%of the AB $h/e$ oscillations will occur when the zero-field 
%(or geometrical) phase difference 
%between the two arms, $\Delta(k_F L)$, changes by $\pi/4$.
A shift in the phase of the AB $h/e$ oscillation, 
between the two values $0$ and $\pi$ that are possible in a two-terminal
measurement,
requires $\Delta(k_F L)$ to change by $\pi$ \cite{Buttiker}. 
For a real device, $\it k_F L$ will not increase
in exactly the same manner for the two arms, and hence the phase and the amplitude
depends on the gate voltage, as we observe.
%This of course also applies to the AB phase \cite{Pedersen00}.
The presence of several transverse subbands
in the arms  will provide an 
additional source of amplitude and phase variation.

The temperature dependence of the AB oscillation amplitude
is shown in Fig.\ \ref{abtempFFT_5}c
for 10 
different gate voltages. There
is some scatter of the data points, around a well-defined average.
The temperature dependence of the oscillation amplitude
does not depend strongly on gate voltage, in the gate voltage interval used here.

We take advantage of the weak gate
voltage dependence, and perform an average of the
Fourier spectra obtained at different gate voltages.
In Fig.\ \ref{abtempFFT_5}a we show Fourier spectra averaged over the 10 gate
voltages, for different temperatures. We also show an average Fourier spectrum 
computed from spectra obtained at 500 different gate voltages at $T$ $=$ 0.32 $\rm K$.
Apart from the peak at the $e/h$ Aharonov-Bohm frequency, clear peaks appear
at frequencies $2e/h$, $3e/h$, $4e/h$, and smaller bumps are also visible
at $5e/h$, $6e/h$. 
Electrons can travel around the ring more than once, and
a periodicity of $h/ne$ in flux of the conductance means
that the interfering electron has enclosed the ring $n$ times.
The 
probability for this type of event to happen
will decrease with $n$, and so will also the
amplitude  of the $ne/h$ oscillation
(as seen in  Fig.\ \ref{abtempFFT_5}a), with a rate that
depends on the coupling between the ring and the 2DEG.

\begin{figure}
\centerline{
\includegraphics[angle=0, width=8.5cm]{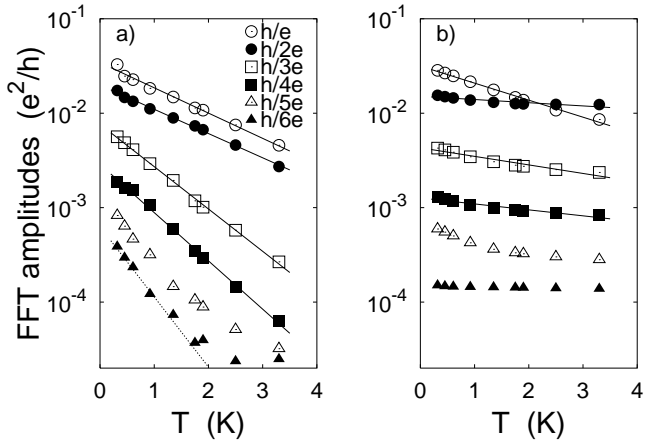}
}
\caption{a) Measured amplitudes of the $h/ne$ oscillations, 
$n=1 \ldots 6$, on a semi-log scale as function of temperature. 
Straight lines are fits with $a_n\cdot \exp (-b_n\cdot T)$,
in the intervals that the lines cover.
Dashed line shows the estimated decay rate of the $h/6e$ oscillation,
$b_6=6a$ (see Fig.\ \ref{abtempRealSim_7}).
b) As in a), showing now amplitudes due to numerically calculated thermal
averaging, as explained in the text.
Straight lines are fits with $c_n\cdot \exp (-d_n\cdot T)$.
}
\label{abtempRealSim_6}
\end{figure}

In Fig.\ \ref{abtempRealSim_6}a  we show the temperature dependence of the
amplitude of the $h/ne$, $n$ $=$ $1 \ldots 6$ oscillation periods, extracted
from the average spectra as shown in Fig.\ \ref{abtempFFT_5}a. For high 
temperatures and frequencies, the FFT amplitudes collapse onto a 
temperature-independent background spectrum. For temperatures between
0.3 and 4 $\rm K$, the amplitude  drops exponentially with temperature
for $n$ $=$ $1 \ldots 4$, but
with different rates $b_n$ for different $n$. 
We assume that the amplitude of a $h/ne$ oscillation strength is
damped due to phase-breaking in the following manner:\cite{factor2}
\begin{equation}\label{n}
\rm h/ne \hspace{1mm} amplitude \it \propto e^{-nL/L_{\phi}(T)} ,
\end{equation}
where $L_{\phi}$ is the characteristic
length over which correlations in the electron phase are lost.
Eq.\ (\ref{n}) implies that all $h/ne$ amplitudes should
have the same functional dependence on $T$, as is observed. 
The measured exponential dependence means that  $L_{\phi} \sim T^{-1}$.
Furthermore, the 
damping rates $b_n$ (the slope of the lines in Fig.\ \ref{abtempRealSim_6}a) should
increase linearly with $n$. 
This is not exactly the case, even though
the rates do increase with $n$.

This apparent inconsistency between  the data and the expected behavior
due to phase-breaking, Eq.\ 
(\ref{n}), can be understood when accounting for the effect of thermal averaging on the
amplitude of the oscillations. At finite temperature, transport can take place
in an energy window around the Fermi energy.
In the Landauer-B\"uttiker formalism, the zero temperature conductance 
$G$ is then convoluted with the derivative of the Fermi function $f$,% \cite{Datta},
\begin{equation}\label{broad}
G(E_F,T,B)=\int
dE \hspace{1mm} G(E,0,B) \left(-\frac{\partial f(E,E_F,T)}{\partial E}\right)
\end{equation}
Considering thermal averaging only,
the amplitude of the AB
oscillations will on average   decrease with temperature,
only if the oscillations at low temperature change
phase within the
available energy window, %defined by the derivative of the Fermi function, 
$\sim$ 3.5$k_B T$ wide. Phase-shifts of the AB oscillations, 
can occur if the geometrical phase difference, $\Delta(k_F L)$,
between the two interfering
paths changes as function of the Fermi energy, as has been motivated above.
Therefore, the relevant temperature scale on which thermal averaging becomes
efficient, is given by the Fermi energy change required to shift the phase of the
AB oscillations, rather than for instance 
the energy level spacing of ring eigenstates. 

We use Eq.\ (\ref{broad}) to estimate the effect of thermal averaging
in our experiment. For $G(E_F,0,B)$ we use a data set $G(V_g,T=0.32\rm K\it,B)$ ,
where $V_g$ is changed in steps of 0.6 $\rm mV$, small enough to resolve
all the changes in the AB oscillations.
The relation of gate voltage $V_g$ to Fermi energy $E_F$ is calibrated by
following the spiky features seen on the low-temperature conductance curve 
in Fig.\ \ref{abtempSEM_1}b for finite bias
voltages, $V_{sd}$. They will for small biases move linearly
in the $(V_g,V_{sd})$ - plane with a slope $\delta V_{sd}/\delta V_g$, which
is related to the Fermi energy change with gate voltage as %$\alpha$ $=$
$\delta E_F/\delta V_g$ $=$ $e/2 (\delta V_{sd}/\delta V_g)$. 
The extracted $\delta E_F/\delta V_g$ is close to a simple capacitor estimate.
From the simulated data sets $G(V_g,T>0.32 \rm K\it,B)$, 
the average AB oscillation amplitude is extracted in the same manner as for the
measured data. The result is shown in Fig.\ 
\ref{abtempRealSim_6}b. The calculated 
oscillation amplitudes do decrease with temperature,
but much slower than the measured amplitudes. 
We conclude that thermal
averaging alone can not account for the measured data. Furthermore, there is a
clear tendency that the calculated $h/ne$ amplitudes decay faster for $n$ odd than even.
For $n$ even, the magnetoconductance oscillations result partly from
interference between time-reversed paths, the same paths that ultimately
give rise to Altshuler-Aronov-Spivak oscillations in disordered systems
\cite{Altshuler81}.
The geometrical phase difference of these paths is 
zero, which means that the resulting part of the magneto-oscillations 
does not change phase.
Consequently they are insensitive to thermal averaging.
This 
explains why the even frequencies in
Fig.\ \ref{abtempRealSim_6}b have a slower temperature dependence than the
odd frequencies.
The temperature dependence can be approximated by
exponentials, lines in Fig.\ \ref{abtempRealSim_6}b.
For diffusive systems \cite{disorderedRings}, 
thermal averaging is not important
for temperatures below $E_c/k_B$, $\sim$ 15-20 $\rm K$ for our system, larger than
the measurement temperatures. Here 
$E_c=\hbar\pi^2D/(2L)^2$ is the standard expression for the correlation energy,
and the diffusion coefficient $D=v_F^2\tau$ in 1D.
But numerical calculations on a ballistic ring have given an exponential temperature
dependence for low temperatures \cite{Shin96}.

\begin{figure}
\centerline{
\includegraphics[angle=0, width=8.5cm]{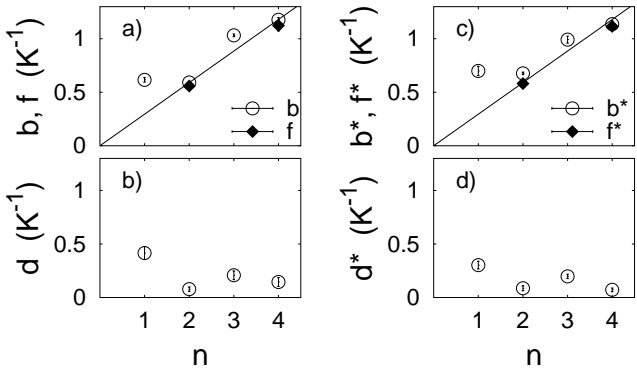}
}
\caption{
a) Circles: exponents $b_n$ from the measured data Fig.\ \ref{abtempRealSim_6}a vs.\ $n$. 
Error bars on the symbols take into account the standard deviations on the fits.
Straight line: fit with  $b_n$$=$$\alpha\cdot n$, $\alpha$ $=$ 0.3 $\rm K^{-1}$ to the $n$ = $2$
and  $n$ = $4$ points. Diamonds: exponents $f_n$ obtained from the temperature dependence
of averaged AB oscillations, as described in the text.
b) Exponents $d_n$ from the simulation Fig.\ \ref{abtempRealSim_6}b vs.\ $n$. 
Error bars on the symbols take into account the standard deviations on the fits
and the Fermi energy calibration.
c) and d) as in a) and b), for another sample.
}
\label{abtempRealSim_7}
\end{figure}

We can now return to the experimental results in Fig.\ \ref{abtempRealSim_6}a. 
As shown, the 
measured $h/e$ ($h/3e$) decay rate has a larger
contribution from thermal averaging than the $h/2e$ ($h/4e$) decay rate.
We see that the part of the damping that thermal averaging can
not account for, approaches the scaling
with $n$ as foreseen in Eq.\ (\ref{n}). 
To demonstrate this, we show
in Fig.\ \ref{abtempRealSim_7}a  the measured $h/ne$ decay rates vs.\ $n$,
and in  \ref{abtempRealSim_7}b the estimated decay rates due
to thermal averaging. The measured rates $b_n$
are not directly proportional to $n$.
But the $h/2e$, $h/4e$ decay rates, which are only little influenced by
thermal broadening (since $b_2/d_2$, $b_4/d_4$ $\sim$ 8), do obey the scaling
(straight line).
Even the $h/6e$ decay rate extrapolated from the scaling
agrees with the data (dashed line in Fig.\ \ref{abtempRealSim_6}a).
From the data in Fig.\ \ref{abtempRealSim_7}a-b
it seems plausible, that with a proper deconvolution
of the thermal averaging from the phase-breaking, also
the  $h/e$, $h/3e$ decay rates will obey the scaling.
The inclusion of the $ne/h$ frequencies, $n$$>$1,
as shown in Fig.\ \ref{abtempRealSim_7}, gives a consistency check that the
temperature damping of the AB oscillations results from
a phase-breaking process, in addition to the effect of thermal averaging,
and this supports the assumption Eq.\ (\ref{n}).
This has to our knowledge not been done before. 
As a further cross-check of the analysis we have also considered the 
amplitude of averaged, measured AB oscillations (as opposed to the
average Fourier spectra, that has been used
above). On performing the average, the amplitude of $h/ne$ oscillations for
$n$ odd is strongly reduced. The temperature dependence of the
amplitude of the ensemble averaged
$h/2e$, $h/4e$ oscillations shows again an exponential decay. The exponents $f_n$
(diamonds in Fig.\ \ref{abtempRealSim_7}a) are only slightly smaller
than the ones obtained from the ensemble averaged spectra, showing again that
these oscillations are only little influenced by averaging.
Finally, with the slope 0.3 % $\pm$ 0.05 
$\rm K^{-1}$ of the line in  Fig.\ \ref{abtempRealSim_7}a, we deduce from
Eq.\ (\ref{n}) that $L_{\phi}(T=1\rm K)$ $=$ 5 $\pm$ 1 $\mu \rm m$ \cite{factor2}.
Another way of expressing our result is that the damping of the interference
amplitude $\sim$ $\exp (-A L k_B T/\hbar v_F)$, where $A$ $\sim$ 0.4.
The same analysis performed on data from the second sample gives similar
results (Fig.\ \ref{abtempRealSim_7}c-d).
Apart from these two rings, we have found interference amplitudes
depending on temperature like $a\cdot \exp (-b\cdot T)$ in 5 other rings
of different designs and sizes.

In summary, we have measured the temperature dependence of phase-breaking
in quasi-1D AB ring structures. We have detected oscillation
amplitudes resulting from the interference of electron paths encircling
the ring more than once. This has enabled us to verify a basic property
of phase breaking: the damping of the amplitude scales with the 
length of the interfering paths.
We have in the analysis included the effect of thermal
averaging, which is determined by the phase-shifts of 
the AB oscillations and hence by the asymmetry changes of the ring.
We find that the phase-breaking length is proportional
to $T^{-1}$, close to what has been measured
in open quantum dots \cite{Bird95,Huibers98}.
Thus, the $T^{-1}$ dependence of phase-breaking might be a general
characteristic of mesoscopic ballistic systems.
A theoretical effort to address this question
is needed.

{\sl Acknowledgments.}
The authors wish to thank H.\ Smith and K.\ Flensberg for useful
discussions.
This work was supported by the Danish Technical Research Council
(grant no.\ 9701490), the Danish Natural Science  Research Council
(grant no.\ 9903274), and by the EU (LTR Programme Q-SWITCH, grant no.\ 30960).
The III-V materials used were made at the III-V Nanolab,
operated jointly by the Microelectronics center of the Danish Technical University
and the Niels Bohr Institute fAFG, University of Copenhagen.

%\clearpage


\begin{references}
\bibitem{dE} The energy spacing $\Delta E_{0,1}$
of the first transverse subband in the exit and
entrance wires is $\sim$ 9 $\rm meV$.
For 
the arms of the ring, we estimate
$\Delta E_{0,1}$ to 1-3 $\rm meV$.
\bibitem{factor2} It can be argued, that $L$ should be a circumference rather than the length of an arm. In that case, we underestimate $L_{\phi}$ by a factor of 2.
\end{references}
\end{document}